# Spin Waves and Magnetic Exchange Interactions in CaFe$_2$As$_2$

Jun Zhao[1], D. T. Adroja[2], Dao-Xin Yao[3], R. Bewley[2], Shiliang Li[1], X. F. Wang[4], G. Wu[4], X. H. Chen[4], Jiangping Hu[3], and Pengcheng Dai[1,5]

[1] Department of Physics and Astronomy, The University of Tennessee, Knoxville, Tennessee 37996-1200, USA

[2] ISIS Facility, Rutherford Appleton Laboratory, Chilton, Didcot, Oxfordshire OX11 0QX, United Kingdom

[3] Department of Physics, Purdue University, West Lafayette, Indiana 47907, USA

[4] Hefei National Laboratory for Physical Sciences at Microscale and Department of Physics, University of Science and Technology of China, Hefei, Anhui 230026, China

[5] Neutron Scattering Science Division, Oak Ridge National Laboratory, Oak Ridge, Tennessee 37831, USA

**Antiferromagnetism is relevant to high temperature (high-$T_c$) superconductivity because copper oxide and iron arsenide high-$T_c$ superconductors arise from electron- or hole-doping of their antiferromagnetic (AF) ordered parent compounds[1-6]. There are two broad classes of explanation for the phenomenon of antiferromagnetism: in the "local moment" picture, appropriate for the insulating copper oxides[1], AF interactions are well described by a Heisenberg Hamiltonian[7,8]; while in the "itinerant model", suitable for metallic chromium, AF order arises from quasiparticle excitations of a nested Fermi surface[9,10]. There has been contradictory evidence regarding the microscopic origin of the AF order in iron arsenide materials[5,6], with some favoring a localized picture[11-15] while others supporting an**

**itinerant point of view[16-20]. More importantly, there has not even been agreement about the simplest "effective ground state Hamiltonian" necessary to describe the AF order[21-25]. Here we report inelastic neutron scattering mapping of spin-wave excitations in $CaFe_2As_2$ (refs. 26, 27), a parent compound of the iron arsenide family of superconductors. We find that the spin waves in the entire Brillouin zone can be described by an effective three-dimensional local moment Heisenberg Hamiltonian, but the large in-plane anisotropy cannot. Therefore, magnetism in the parent compounds of iron arsenide superconductors is neither purely local nor purely itinerant, rather it is a complicated mix of the two.**

Since the discovery of static AF order (with a spin structure in Fig. 1a) in the parent compounds of iron pinictide superconductors[5,6], much effort has been focused on understanding the role of spin dynamics in the superconductivity of these materials[11-20]. A determination of the effective magnetic exchange coupling and ground state Hamiltonian in the parent compounds of these materials is important because such an understanding will provide the basis against which superconductivity-induced changes can be identified. Using inelastic neutron scattering, we have measured the dispersion of spin-wave excitations in $CaFe_2As_2$ (refs. 26, 27), one of the parent compounds of the FeAs-based superconductors, and determined the effective magnetic exchange interactions. If the static long-range AF order depicted in Fig. 1a for the parent compounds of iron-based superconductors originates from a collective spin-density-wave order instability of itinerant electrons like in chromium, the velocity of spin wave excitations $c$ should be $c = (v_e v_h/3)^{1/2}$, where $v_e$ and $v_h$ are the electron and hole Fermi velocity, respectively[9]. Furthermore, spin wave excitations should exhibit longitudinal and transverse polarization, and damp into single particle excitations (Stoner continuum) via transfer of an electron (spin) from the majority to the minority band at high energies as shown schematically in Fig. 1c (ref. 10). On the other hand, if magnetic order in iron pnictides has a local moment origin as in the parent compounds of the copper oxides[1], one should observe well-defined (essentially instrumental resolution limited) spin waves throughout the Brillouin zone and magnetic coupling between local moments should be dominated by direct and super-exchange interactions (Fig. 1d)[11-15]. Although the presence of itinerant magnetic excitations and Stoner continuum have been suggested in $BaFe_2As_2$ (ref. 24) and $CaFe_2As_2$ (ref. 25), these measurements were carry out at energies well below the zone boundary spin wave energy (~200 meV) and therefore were unable to conclusively determine the effective magnetic exchange interactions and life time of the spin waves.

We used inelastic neutron scattering to study low-temperature ($T = 10$ K) spin waves of single crystals of $CaFe_2As_2$ which has a Néel temperature of $T_N \approx 170$ K (refs. 26, 27). Figure 1e-l shows two-dimensional constant-energy (-$E$) images of spin-wave excitations of $CaFe_2As_2$ around the AF zone center in the ($H$, $K$) scattering plane[21-25]. Previous low-energy measurements[23] revealed that spin waves in $CaFe_2As_2$ are three-dimensional and centered at AF wave vector $Q = (1, 0, L = 1, 3, 5, …)$ reciprocal lattice units (rlu). For energy transfers of $E = 48 \pm 6$ (Fig. 1e) and $65 \pm 4$ meV (Fig. 1f), spin waves are still peaked at $Q = (1,0,)$ rlu in the center of the Brillouin zone (shown as dashed square boxes). As energy increases to $E = 100 \pm 10$ (Fig. 1g), $115 \pm 10$ (Fig. 1h), $137.5 \pm 15$ (Fig. 1i), $135 \pm 10$ (Fig. 1j), and $145 \pm 15$ meV (Fig. 1k), counter-propagating spin-wave modes become apparent. The scattering changes from ring-like at 100 meV

(Fig. 1g) to ellipses elongated along the *K*-direction for energies above 110 meV (Figs. 1h-1k). For an energy transfer of $175 \pm 15$ meV (Fig. 1l), spin waves show a broad square-like scattering already reaching the zone boundary in the *K*-direction.

To quantitatively determine the spin-wave dispersion, we cut through the two-dimensional images similar to Fig. 1 for various incident beam energies ($E_i$) aligned along the *c*-axis. Figures 2a-g show the outcome for different spin-wave energies in the form of constant-*E* scans along the *K*-direction around the AF zone center. As the excitation energy increases from 25 meV (Fig. 2g) to 144.5 meV (Fig. 2a), well-defined counter-propagating spin waves approach the zone boundary. To illustrate the general feature of the high-energy spin waves, we have used the scattering near (2,0,0) rlu as a background and assumed the positive scattering at wave vectors below (2,0,0) rlu is entirely magnetic. Figure 3a shows the outcome of the background subtracted scattering for the $E_i = 450$ meV data projected in the wave vector ($Q = [1,K]$) and energy space. In spite of the spin wave intensity modulation along the *L*-direction due to the exchange interaction $J_c$ between the FeAs planes[23] (Fig.1a), one can see three clear plumes of scattering arising from the in-plane AF zone centers $Q = (1,-2), (1,0)$, and $(1,2)$ rlu. The spin-wave scattering disperses for energies above 100 meV and extends up to about 200 meV. Since spin waves become less dispersive as the zone boundary is approached, we locate the spin wave excitations via energy scans at a fixed wave vector. Figures 3c-h summarize a series of such scans at different wave vectors which reveal clear dispersions near the zone boundary and a maximum spin-wave bandwidth of about 200 meV.

In addition to the results presented in Figs. 1-3, we have also collected similar data at other wave vectors throughout the Brillouin zone. The solid circles in Figures 4a-c summarize our measured spin wave dispersions along the [H,0,1], [1,0,L], and [1,K,1] directions. To understand these data as well as the wave vector-energy (*Q-E*) dependence of the spin-wave intensities, we consider a Heisenberg Hamiltonian consisting of effective in-plane nearest-neighbors (Fig. 1a, $J_{1a}$ and $J_{1b}$), next-nearest-neighbor (Fig. 1a, $J_2$), and out-of-plane (Fig. 1a, $J_c$) exchange interactions. The dispersion relations are given by[21-25]: $E(q) = \sqrt{A_q^2 - B_q^2}$, where $A_q = 2S[J_{1b}(\cos(\pi K) - 1) + J_{1a} + J_c + 2J_2 + J_s]$, $B_q = 2S[J_{1a}\cos(\pi H) + 2J_2\cos(\pi H)\cos(\pi K) + J_c\cos(\pi L)]$, $J_s$ is the single ion anisotropy constant, and $q$ is the reduced wave vector away from the AF zone center. The neutron scattering cross section can be written as[22]:

$$\frac{d^2\sigma}{d\Omega dE} = \frac{k_f}{k_i}\left(\frac{\gamma r_0}{2}\right)^2 g^2 f^2(Q) e^{-2W} \sum_{\alpha\beta}(\delta_{\alpha\beta} - Q_\alpha Q_\beta) S^{\alpha\beta}(Q,E),$$ where

$(\gamma r_0/2)^2 = 72.65$ mb/sr, $g$ is the $g$-factor ($\approx 2$), $f(Q)$ is the magnetic form factor of iron $Fe^{2+}$, $e^{-2W}$ is the Debye-Waller factor ($\approx 1$ at 10 K), $Q_\alpha$ is the $\alpha$ component of a unit vector in the direction of $Q$, $S^{\alpha\beta}(Q,E)$ is the response function that describes the $\alpha\beta$ spin-spin correlations, and $k_i$ and $k_f$ are incident and final wave vectors, respectively. Assuming that only the transverse correlations contribute to the spin-wave cross section and finite excitation lifetimes can be described by a damped simple harmonic oscillator with inverse lifetime $\Gamma$ (refs. 28, 29), we have $S^{yy}(Q,E) = S^{zz}(Q,E) =$

$$S_{eff}\frac{(A_q-B_q)}{E_0(1-e^{-E/k_BT})}\frac{4}{\pi}\frac{\Gamma E E_0}{(E^2-E_0^2)^2+4(\Gamma E)^2},$$ where $k_B$ is the Boltzmann constant, $E_0$

is the spin-wave energy, and $S_{eff}$ is the effective spin. We analyzed our data by keeping $S$ and $S_{eff}$ distinct following the practice of Ref. 22.

We fitted the measured absolute intensity of spin wave excitations and their dispersions in Figs. 1-4 by convoluting the above discussed neutron scattering spin-wave cross section with the instrument resolution using Tobyfit program[28,29]. Since $CaFe_2As_2$ exhibits tetragonal to orthorhombic lattice distortion below the $T_N$ (ref. 27), care was taken to include the $(H,K)/(K,H)$ twin domains in the computed scattering cross section. We find that the Heisenberg Hamiltonian with only the nearest-neighbors effective exchange couplings ($J_{1a}$ and $J_{1b}$ are finite, and $J_2 = 0$) cannot explain the data. Theoretically, it has been argued that the observed collinear spin structure in Fig. 1a is consistent with either $SJ_{1a} \approx SJ_{1b} \approx \frac{1}{2}SJ_2$ or $SJ_{1a} \approx 2SJ_2 \gg SJ_{1b}$, and distinguishing these two models require spin-wave data near the zone boundary[22]. Although previous neutron scattering experiments on $CaFe_2As_2$ suggest $SJ_{1a} \approx SJ_{1b} = 25 \pm 8$, $SJ_2 = 36 \pm 2$, and $SJ_c = 7 \pm 1$ meV (refs. 23, 25), these results are obtained by fitting spin-wave data well below the ~200 meV zone boundary energy (Fig. 3) and therefore are inconclusive.

The red dashed lines in Figs. 3f-h show the expected zone boundary spin waves assuming $SJ_{1a} = 27$, $SJ_{1b} = 25$, $SJ_2 = 36$, and $SJ_c = 5.3$ meV. It is obvious that such a model failed to describe the zone boundary data. Our best fits to both the low-energy and zone boundary spin waves by independently varying the effective exchange parameters are shown as solid black lines in Figs. 2 and 3 with $SJ_{1a} = 49.9 \pm 9.9$, $SJ_{1b} = -5.7 \pm 4.5$, $SJ_2 = 18.9 \pm 3.4$, $SJ_c = 5.3 \pm 1.3$ meV. The broadening of the spin waves with increasing energy is accounted for via $\Gamma \propto 0.15E$ and shown as dotted lines in Fig. 4. From our best fit to all spin wave data, we find $S_{eff} = 0.22 \pm 0.06$ which is somewhat smaller than previous measurements on powder samples of $BaFe_2As_2$ (ref. 22). The value of $S_{eff}$ and the measured 0.8 $\mu_B$/Fe static moment[27] suggest of a $S \sim 1/2$ system. Theoretically, if we consider a spin 1/2 quantum Heisenberg model with the above exchange parameters, a simple calculation reveals elastic moment = $g(S-\Delta S)$ $\mu_B$ = 2(1/2-0.09) $\mu_B$ = 0.82 $\mu_B$, where $\Delta S$ is the spin wave correction to the magnetic moment in quantum Heisenberg model[30], and $S_{eff} = Z_d S = 0.285$, where $Z_d = 0.57$ is an intensity-lowering renormalization factor of the one magnon cross section due to quantum fluctuations and magnon-magnon interactions[30].

From fitting results in Figs. 2-4, we see that the spin wave dispersion and intensity in CaFe$_2$As$_2$ throughout the Brillouin zone can be well described by a Heisenberg Hamiltonian with effective nearest-neighbors and next-nearest neighbor exchange interactions. Figures 4a-c summarize the spin-wave dispersions along all three high symmetry directions and Figure 4d shows energy-dependence of the local susceptibility[7], together with calculations using $SJ_{1a} \approx SJ_{1b}$ (red dashed lines) or our (solid lines) models. The former model clearly fails to describe the data. To test whether the spin-wave branch crosses the Stoner continuum as schematically illustrated in Fig. 1c, we plot spin-wave damping $\Gamma$ versus $E$ as dotted lines in Figs. 4a-c. Although $\Gamma$ is approximately proportional to $0.15E$, there is no steep increase in $\Gamma$ at any wave vector indicative of a Stoner continuum (Fig. 1c). Instead, the observed spin-wave broadening at high energies may arise from magnon-electron scattering due to the low-temperature metallic nature of the system, similar to ferromagnetic metallic manganites[28,29].

The central message of our work is that one can fit spin waves of CaFe$_2$As$_2$ throughout the Brillouin zone with a simple Heisenberg Hamiltonian without the need for Stoner continuum— the hall mark of an itinerant electron system. The lack of direct evidence for a Stoner continuum below 200 meV suggests weak low energy electron-hole particle excitations. One local density approximation calculation has predicted essentially the correct in-plane magnetic exchange couplings[20], these results, however, are obtained within the tetragonal and collinear AF ordered structures contrary to the experiments. Furthermore, band structure calculations suggest that the Fermi velocity $a/b$ anisotropy in $CaFe_2As_2$ is less than 8% in the low temperature orthorhombic phase (D. J. Singh, private communication). If spin-wave velocities in $CaFe_2As_2$ are proportional to $(v_e v_h/3)^{1/2}$ as those in chromium[9], they should be similar along the $a/b$ directions. Although our results appear to favor a localized moment picture, a spin 1/2 model cannot be produced if all orbitals in iron are localized since there are even numbers of electrons per iron. Moreover, it is difficult to understand why direct and super-exchange interactions within the Fe-As-Fe plane are so different along the $a/b$ directions of the orthorhombic structure because the tetragonal to orthorhombic lattice distortion below $T_N$ is small and only weakly affects the Fe-As-Fe bond distances/angles[5,6]. The observed large difference may hint the involvement of other electronic degree of freedoms, such as orbital, in the magnetic transition. To achieve a comprehensive understanding of spin excitations, one must consider both the localized and itinerant electrons in these materials.

**Acknowledgements** We thank A. T. Boothroyd, Toby Perring, David Singh, Andriy Nevidomsky for helpful discussions. This work is supported by the US NSF and DOE Division of Materials Science, Basic Energy Sciences. This work is also supported by the US DOE through UT/Battelle LLC. The work at USTC is supported by Natural Science Foundation of China, the Chinese Academy of Sciences and the Ministry of Science and Technology of China.

**Figure 1 Magnetic structure, calculated spin-wave dispersion and wave vector dependence of spin-wave excitations at different energies for $CaFe_2As_2$.** Our inelastic neutron scattering experiments were carried out on the MERLIN time-of-flight chopper spectrometer at the Rutherford-Appleton Laboratory, Didcot, UK. We co-aligned 6.4 grams of single crystals of $CaFe_2As_2$ grown by self-flux (with in-plane mosaic of 2 degrees and out-of-plane mosaic of 3 degrees). The incident beam energies were $E_i$ = 50, 80, 150, 200, 250, 450, 600 meV, and mostly with $E_i$ parallel to the $c$ axis. Spin wave intensities were normalized to absolute units using a vanadium standard (with 30% error). We define the wave vector $Q$ at $(q_x, q_y, q_z)$ as $(H, K, L) = (q_x a/2\pi, q_y b/2\pi, q_z c/2\pi)$ rlu, where $a = 5.506$, $b = 5.450$, and $c = 11.664$ Å are the orthorhombic cell lattice parameters at 10 K (ref. 27). a) Schematic diagram of the Fe spin ordering in $CaFe_2As_2$. b) Calculated three-dimensional spin-wave dispersions using $SJ_{1a} = 49.9$, $SJ_{1b} = -5.7$, $SJ_2 = 18.9$, and $SJ_c = 5.3$ meV. c) Schematic diagram for how spin-wave dispersion enters into Stoner continuum. d) Dispersion of spin waves in a classical Heisenberg Hamiltonian. Wave vector dependence of the spin waves for energy transfers of e)

$E = 48 \pm 6$ meV [$E_i = 150$ meV and $Q = (1,0,3)$]; f) $E = 65 \pm 4$ meV [$E_i = 250$ meV and $Q = (1,0,3)$]; g) $E = 100 \pm 10$ meV [$E_i = 450$ meV and $Q = (1,0,3.5)$]; h) $E = 115 \pm 10$ meV [$E_i = 450$ meV and $Q = (1,0,4)$]; i) $E = 137 \pm 15$ meV [$E_i = 600$ meV and $Q = (1,2,4)$]; j) $E = 135 \pm 10$ meV [$E_i = 450$ meV and $Q = (1,0,4.5)$]; k) $E = 144 \pm 15$ meV [$E_i = 450$ meV and $Q = (1,0,5)$]; l) $E = 175 \pm 15$ meV [$E_i = 600$ meV and $Q = (1,0,5.2)$].

**Figure 2 Constant energy cuts of the spin-wave dispersion as a function of increasing energy and our model fit using the Heisenberg Hamiltonian.** A series of constant-energy cuts through the AF spin-wave zone center as a function of decreasing energy a) $E = 144 \pm 20$; b) $E = 135 \pm 10$; c) $E = 115 \pm 15$; d) $E = 100 \pm 10$; e) $E = 64 \pm 10$; f) $E = 48 \pm 6$; g) $E = 25 \pm 5$ meV. The solid lines are model fits to the data after convoluting the cross section to the instrumental resolution. Typical instrumental resolutions are shown as dotted lines in (a) and (d). Error bars indicate one sigma.

**Figure 3 Observed and calculated spin waves at 10 K, and constant-$Q$ cuts near the AF zone boundary.** a) The projections are in the scattering plane formed by the energy transfer axis and $(1,K)$ direction (with integration of $H$ from 0.8 to 1.2 rlu) after subtracting the background integrated from $1.8 < H < 2.2$ and from $-0.25 < K < 0.25$. Data were obtained with $E_i = 450$ meV. b) Calculated spin wave excitations using model specified in the text. c-h) Constant-$Q$ cuts at various wave vectors near the zone boundary obtained with $E_i = 600$ meV. The solid ($SJ_{1a} > 0$, $SJ_{1b} < 0$) lines are our model fits to the data and the dashed lines are calculations assuming $SJ_{1a} \approx SJ_{1b}$.

**Figure 4 Spin-wave dispersion relation along high symmetry directions in the three-dimensional Brillouin zone and energy dependence of the local susceptibility.** The solid circles in Figs. a-c) are extracted from constant-$E(-Q)$ cuts of various $E_i$ data. The horizontal bars indicate the $E(Q)$ integration range and vertical bars are errors calculated from least square fittings. Solid (dashed) lines are fits to spin-wave models discussed in the text. The lengths of the blue vertical bars indicate wave vector dependence of $\Gamma$, the $\Gamma/E \sim 0.15$ is much smaller than metallic ferromagnet $La_{2-2x}Sr_{1+2x}Mn_2O_7$ where $\Gamma/E \sim 0.33 - 0.46$ (ref. 28), thus suggesting smaller influence of itinerant electrons in $CaFe_2As_2$. The blue dotted line is a guide to the eye. d) Energy dependence of the local susceptibility$^2$ obtained by integrating raw intensities above background from $0.5 < H < 1.5$; $-0.5 < K < 0.5$, and $L$ from $L-0.5$ to $L+0.5$, where $L = 1, 3, 5$ in the $(1,0,L)$ zone. Twinning effect has not been taken out. In our experimental set up, the energy, magnetic form factor, and polarization factors are all weakly $Q$ dependent within the Brillouin zone. For simplicity, we used appropriate values for these factors at the zone center $Q = (1,0,L)$. Solid and dashed lines are expected energy dependence of the local susceptibility for the two models discussed in the text with consideration of the twinning effect.

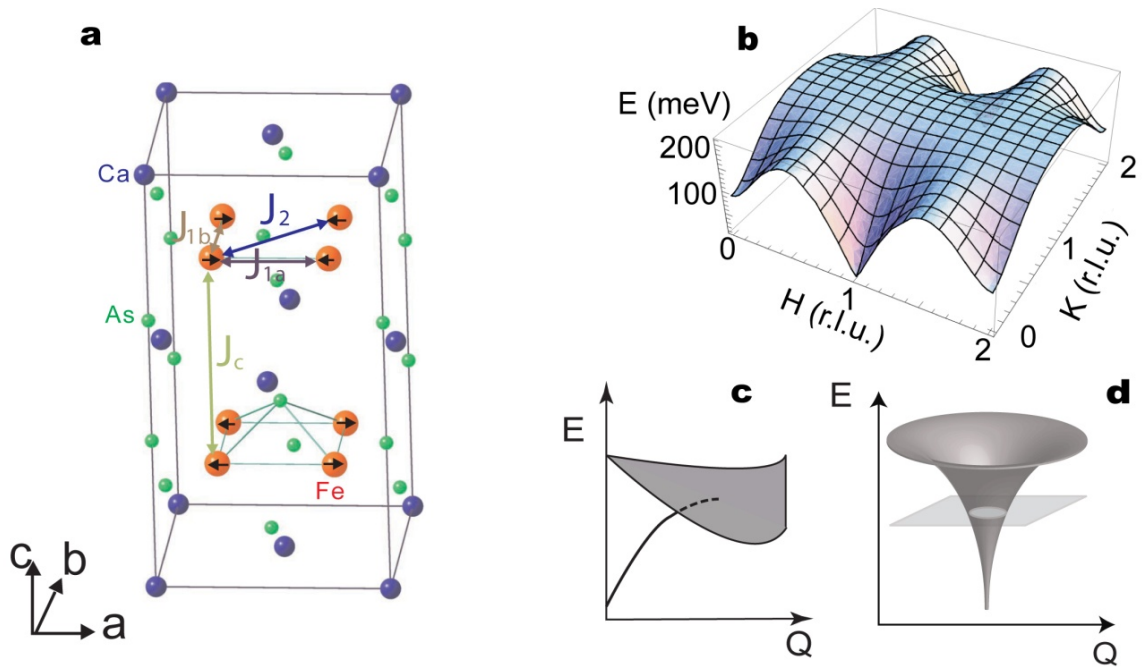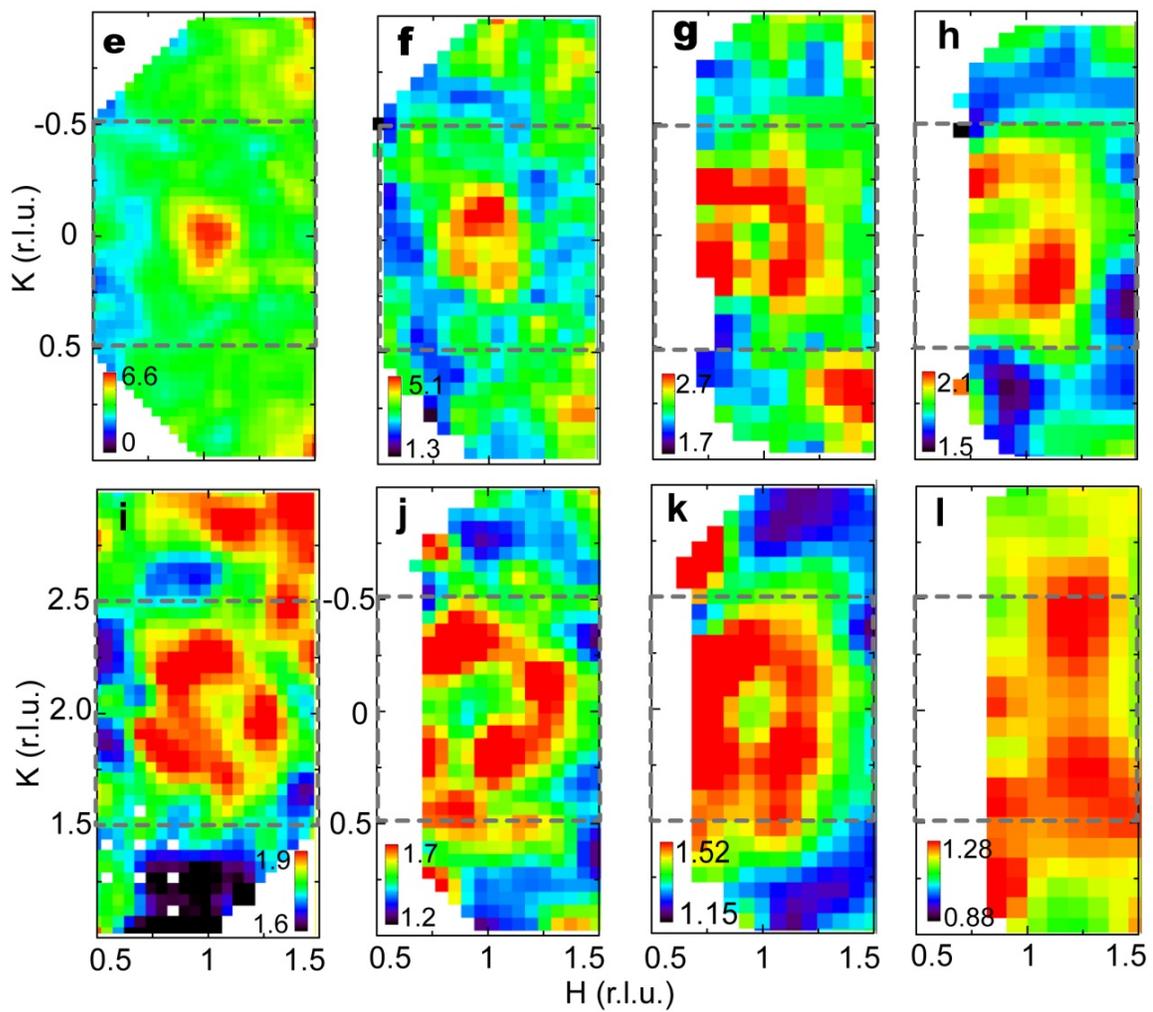

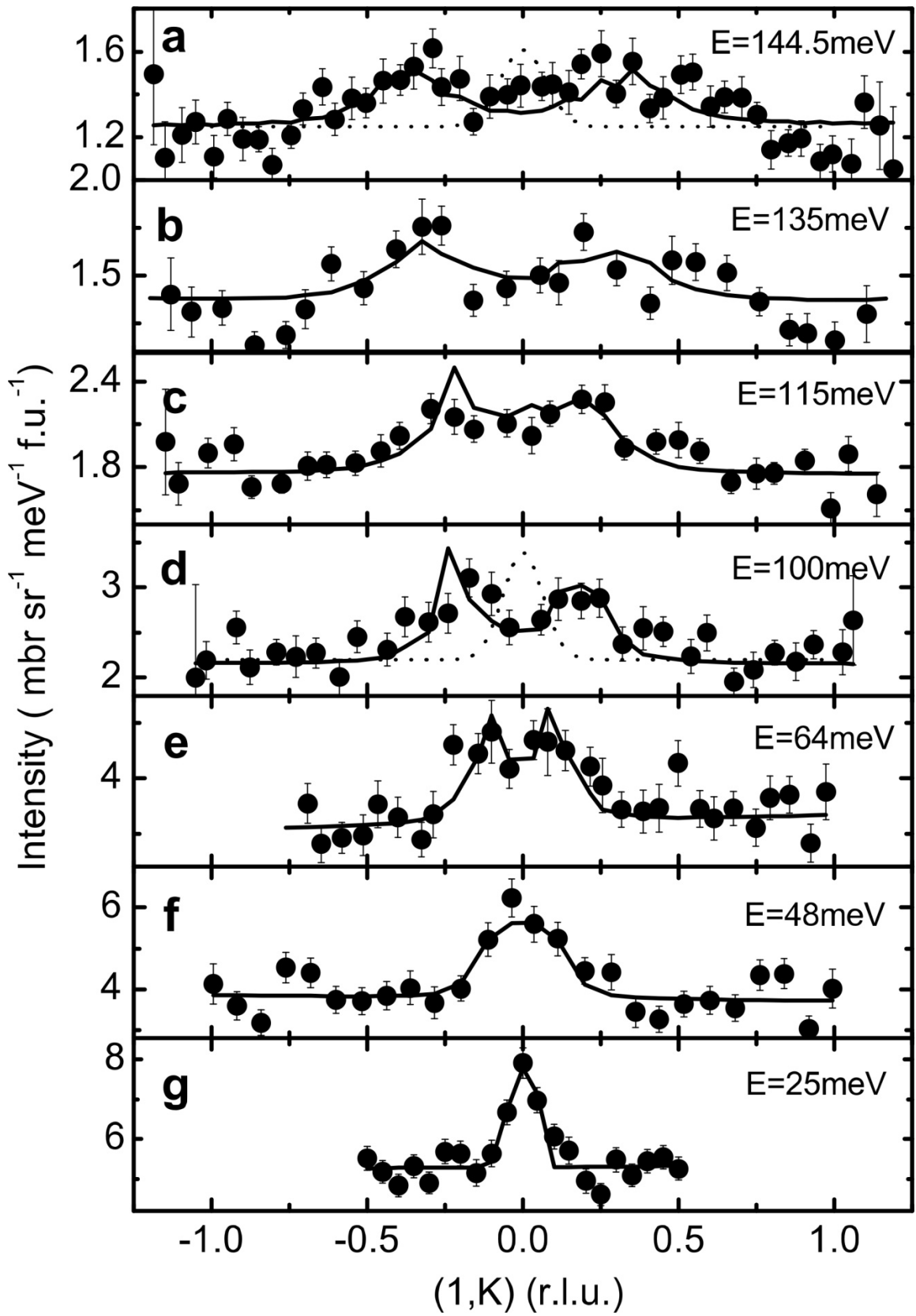

Fig. 2

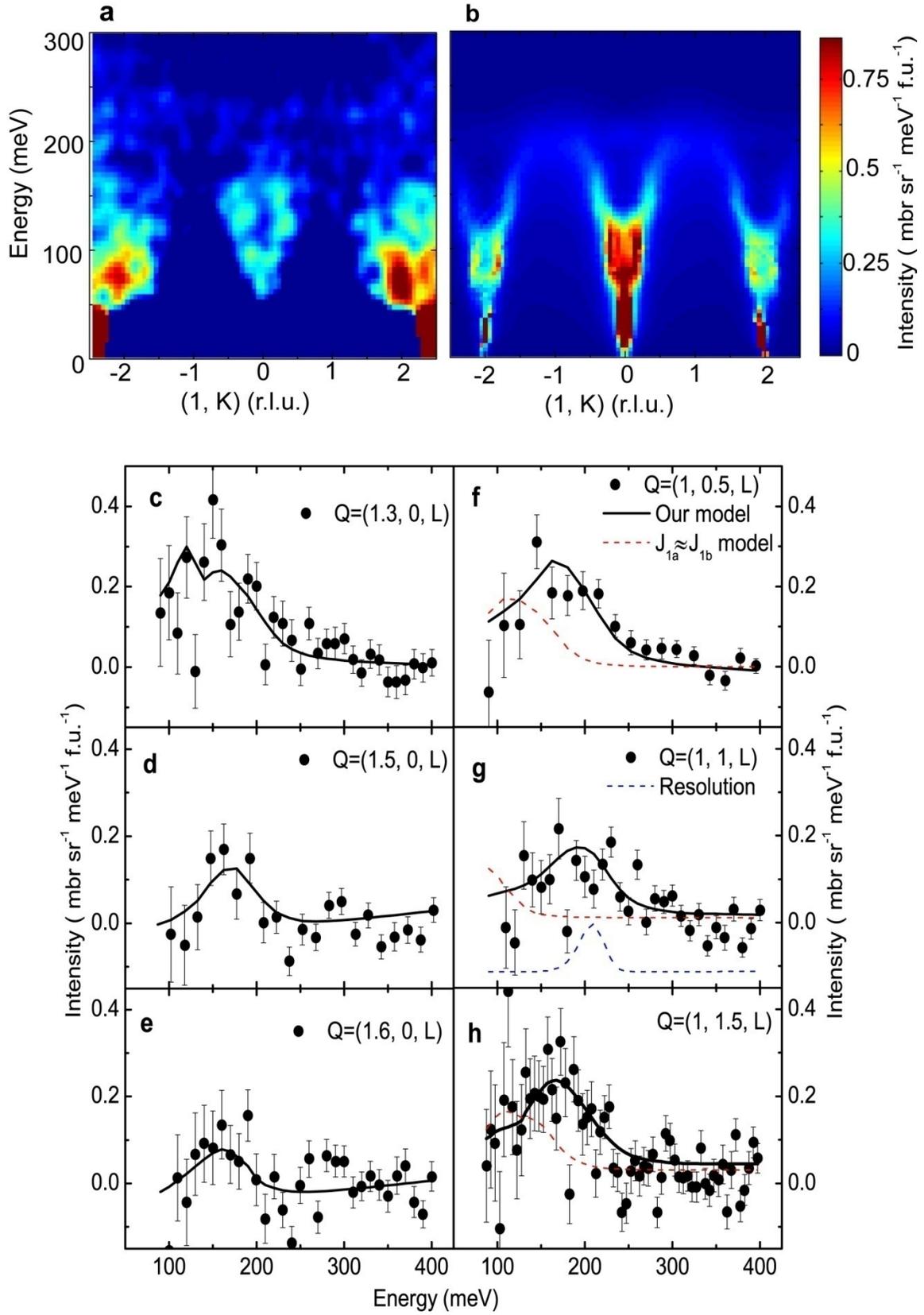

Fig. 3

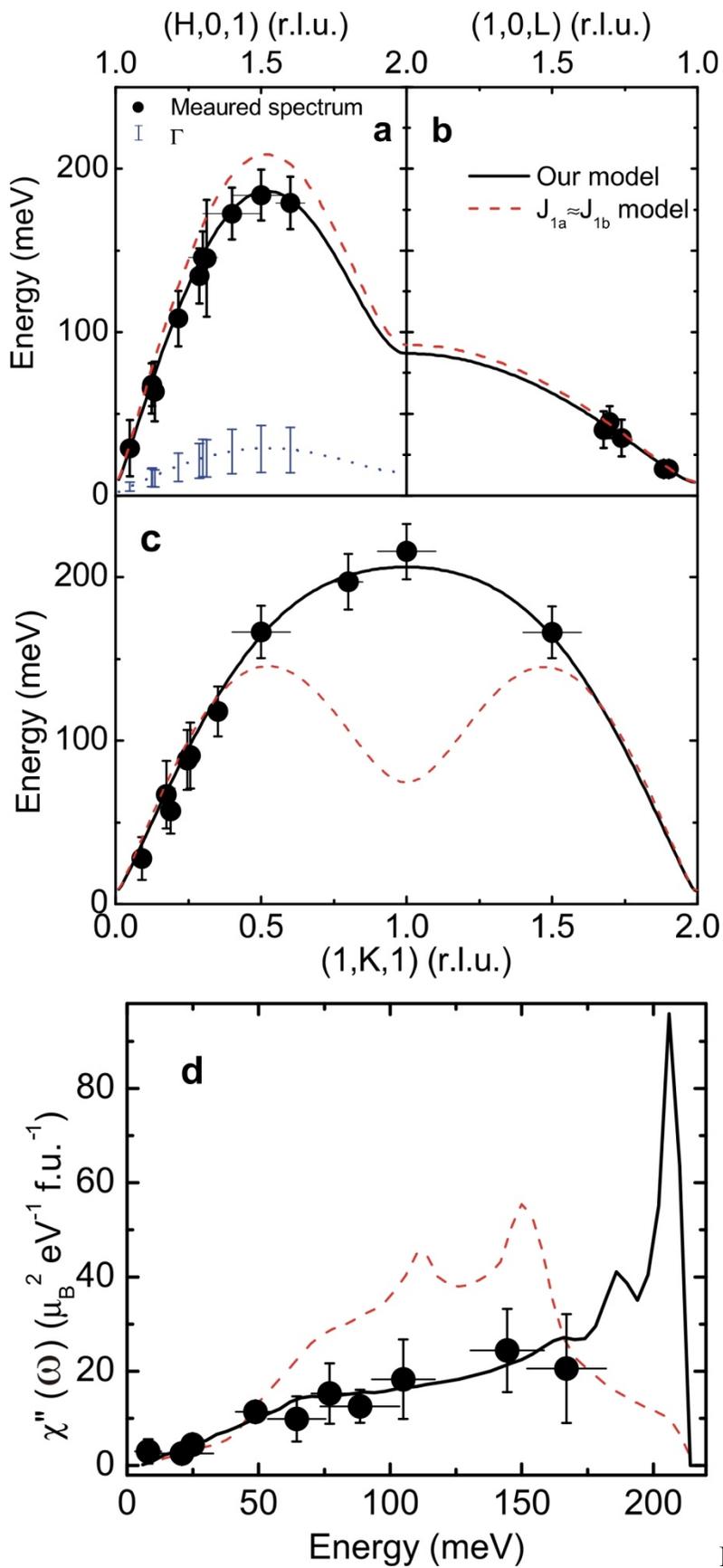

Fig. 4